\documentclass[11pt]{article}   	

\usepackage{graphicx}
\usepackage{csquotes}
\usepackage{color}

\begin{document}

\title{A two-dimensional index to quantify both scientific research impact and scope}

\author{M. Adda-Bedia$^\dag$  and F. Lechenault$^\ddag$\\
         \small{ {\it $^\dag$Univ. Lyon, ENS de Lyon, Univ. Claude Bernard, CNRS,}}\\
         \small{{\it Laboratoire de Physique, Lyon, France}}\\
         \small{{\it $^\ddag$Laboratoire de Physique Statistique, Ecole Normale Sup\'erieure,}}\\
         \small{{\it PSL Research University, CNRS, Paris, France}}
}

\date{\today}

\maketitle

\begin{abstract}
Modern management of research is increasingly based on quantitative bibliometric indices. Nowdays, the $h$-index is a major measure of research output that has supplanted all other citation-based indices. In this context, indicators that complement the $h$-index by evaluating different facets of research achievement are compelling. As an additional bibliometric source that can be easily extracted from available databases, we propose to use the number of distinct journals $N_j$ in which an individual's papers were published. We show that $N_j$ is independent of citation counts, and argue that it is a relevant indicator of research scope, since it quantifies readership extent and scientific multidisciplinarity. Combining the $h$-index and $N_j$, we define a two-dimensional index $(H,M)$ that measures both the output (through $H$) and the outreach (through $M$) of individual's research. In order to probe the relevance of this two dimensional index, we have analysed the scientific production of a panel of physicists belonging to the same Department but with different research themes. The analysis of bibliometric data confirms that the two indices are uncorrelated and shows that while $H$ reliably ranks the impact of researchers, $M$ accurately sorts multidisciplinary or readership aspects. We conclude that the two indices together offer a more complete picture of research performance and can be applied either for individuals, research groups or institutions.
\end{abstract}

\section*{Introduction}

Bibliometric indices are ubiquitous in measuring the scientific impact of both journals, institutions, research groups and individual researchers. At some stage in their evaluation process, scientific committees, administrators or policy makers often rely on citation data to assess scientific output. Among the variety of measures that can be derived from raw citation data, the $h$-index~\cite{Hirsch2005} became the most popular bibliometric criterion for the ranking of research accomplishment~\cite{Ball2005,Fersht2009,Alonso2009}.

The $h$-index of an individual is given by the total number of published papers with citations $\geq h$. Due to its simple definition, its ease of computation from existing bibliography databases, its robustness against errors in the long tails of the citations-rank distribution~\cite{Vanclay2007} and its good properties when quantifying the scientific production and its impact, the $h$-index has been largely adopted as a reliable measure of research output~\cite{Alonso2009}. Nowadays, the automatic calculation of $h$-indices is a built-in feature of major bibliographic databases such as Google Scholar, Researchgate, Scopus and Web of Science. While it is quite dependent on the database being used, the typical value of the $h$-index is robust when it comes to ranking both individuals and research institutions.

Shortcomings of the $h$-index have been pointed out, although many other available indicators do suffer from similar biases~\cite{Alonso2009}. Among the arguments raised against using the $h$-index one can enumerate: it does not allow to compare scientists from different disciplines~\cite{Batista2006}, it does not take into account multi-authored papers~\cite{Iglesias2007,Egghe2008b,Radicchi2008,Sekercioglu2008,Zhang2009,Hirsch2010}, it is a time-dependent quantity~\cite{Burrell2007,Eom2011,Acuna2012,Wang2016}, it does not highlight the citation scores of top articles~\cite{Egghe2006}, it does not take into account the context of the citations in the papers. Consequently, several subsequent variations of the $h$-index have been proposed to overcome some of these drawbacks~\cite{Batista2006,Iglesias2007,Egghe2008b,Radicchi2008,Sekercioglu2008,Zhang2009,Hirsch2010,Burrell2007,Eom2011,Acuna2012,Wang2016,Egghe2006,Komulski2006,Sidiropoulos2007,Jin2007a,Jin2007b,VanEck2008,Rousseau2008,Egghe2008a,Ruane2008,Anderson2008,Bornmann2008,Antonakis2008,Schreiber2009,Guns2009}. A non exhaustive list of such bibliometric indices include the $g$-index~\cite{Egghe2006}, the $h^{(2)}$-index~\cite{Komulski2006}, the $A$- $R$- and $AR$-index~\cite{Jin2007a,Jin2007b}, the $m$-index~\cite{Bornmann2008}, the $h_m$-index~\cite{Schreiber2009}, etc. Interestingly, all of them were based on an increasingly sophisticated analysis of raw citation data, which makes them more difficult to access. Moreover, although being designed to surmount its perceived shortcomings, they were found to be quite positively correlated with the $h$-index~\cite{Alonso2009,Bornmann2008}. This fact yielded the general consensus that no other bibliometric indicator of research output is clearly preferable to the $h$-index~\cite{Alonso2009,Hirsch2007}. 

Since it is desirable to evaluate researchers and/or institutions on more than a single quality, we propose to supplement the $h$-index with a new quantitative indicator specialized to interdisciplinary research and readership extent. This journal-based index is by construction independent of citation counts and measures different qualities that could be useful to assess, for example in the process of hiring individuals who are expected to teach and to advise, for which high specialization values are not required.

\section*{A new two-dimensional index}

Many refinements of the $h$-index, on both theoretical and empirical sides, were based on alternative analysis of citation data. We believe that this is the reason for their strong positive correlation with the latter. Therefore, the quest for other bibliometric indices should be aimed at complementing the $h$-index rather than replacing it. A new index should be a quantitative indicator with a core that is independent of citation data and as easy to compute from existing bibliographic databases as the $h$-index. It should concern both individuals, research groups and institutions and should highlight different achievements than research output, which is already well quantified by the $h$-index. 

Noticing that the $h$-index is bounded by the total number of published papers, $h\leq N$, we have looked for another possibly interesting and simple quantity that satisfies the same property. A straightforward one is the number of different journals in which these papers were published. This number, denoted by $N_j$, is not directly available in existing bibliographic databases but can be easily extracted from them, either by direct counting or using simple script codes. Obviously, $N_j\leq N$ and is not based on citation count, thus one would expect that it is unlikely correlated to research output. Indeed, common publishing habits show that a paper is submitted to a given journal because of the studied area of research, to reach a specific scientific community, to fit a specific format or to be faced with a given more or less selective refereeing procedure. One could argue that a such choice is mainly dictated by the journal's impact factor, its ranking by a target institution or its interdisciplinary readership. Nevertheless, if the published research is worthwhile it will positively impact the $h$-index and consequently research output, regardless of the journal's ranking. Moreover, a paper in a high impact journal with an interdisciplinary readership is often followed by a series of detailed papers in the same subject intended for a specific scientific community, which thus contributes to increase $N_j$.

However, what type of research achievement is embedded in this new number? On the one hand, publishing in a small set of journals could indicate mono-thematic research interests while a multidisciplinary researcher is prone to publish in a wide range of journals. On the other hand, a researcher could impact a single field by publishing in the same journal on a single subject, and conversely, an author may publish about various subjects in many journals by achieving a moderate impact. These two extreme and  antagonistic examples are in favour of taking into account both $N_j$ and the $h$-index for more elaborate ranking purposes.
Although $N_j$ may exhibit some flaws, which will be discussed later, one can reasonably assume that it carries information about either the multidisciplinary nature of the research, the diversity of interests of the researcher, or the extent of his readership. While these features are indicators of research quality, they are clearly not directly provided by the $h$-index.
\begin{figure}[htb]
\begin{center}
   \includegraphics[width=0.6\textwidth]{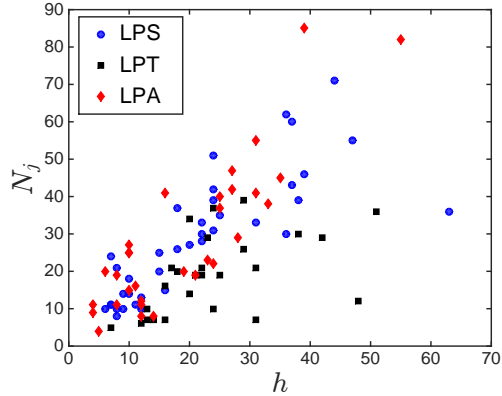} 
   \caption{Scatter plot of $N_j$ as a function of $h$ for a panel of researchers working at the Physics Department of ENS Paris (see Text for details).
   \label{fig:example1}}
\end{center}
\end{figure}

\begin{figure}[htb]
\begin{center}
   \includegraphics[width=0.6\textwidth]{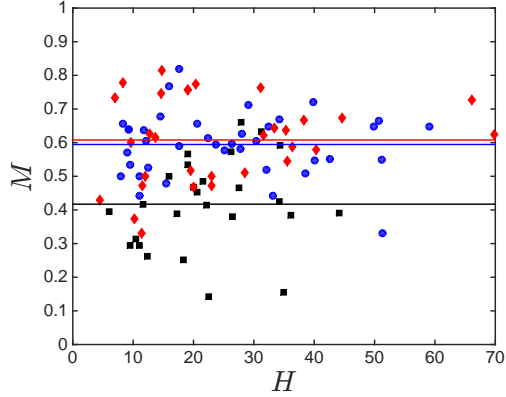} 
   \caption{Scatter plot of $M$ as a function of $H$ for the same panel as in Fig.~\ref{fig:example1}. Continuous lines show the mean $M$-index for each \enquote{Laboratoire} (see Text for details): 0.61 for LPA, 0.42 for LPT and 0.59 for LPS. The corresponding standard deviations are: 0.128 fxor LPA, 0.135 for LPT and 0.094 for LPS.
   \label{fig:example2}}
\end{center}
\end{figure}

To test the relevance of this two-dimensional index, we have chosen as a panel 95 permanent physicists belonging to a same department, namely the Physics Department of ENS Paris, which is divided in three sub-departments (\enquote{Laboratoires})  with more or less specific research areas: condensed matter physics (Laboratoire Pierre Aigrain, LPA, 31 researchers), theoretical and statistical physics (Laboratoire de Physique Th\'eorique, LPT, 26 researchers), statistical physics, biophysics and nonlinear physics (Laboratoire de Physique Statistique, LPS, 38 researchers). The $h$-index and $N_j$ of each individual have been retrieved from Web of Science in July 2016. All types of publications were taken into account. Fig.~\ref{fig:example1} shows the results of this survey: a scatter plot of these quantities, though containing a global trend, is dominated by the scatter noise, illustrating the signals low correlation. This confirms that when the department is taken as a whole, individuals bibliometric indicators $h$ and $N_j$ carry complementary information. This example justifies the use of both $h$ and $N_j$ to characterise research accomplishment. In order to sharpen the comparison between individuals, we propose a different representation of these two bibliometric indicators. Since both $h$ and $N_j$ are bounded by $N$, one can define a \enquote{complex} representation of the two-dimensional index $(h,N_j)$ as follow
\begin{equation}
H \exp\left(\frac{i\pi}{2}M\right)= \frac{h}{\sqrt{2}}+i\,\frac{N_j}{\sqrt{2}}\;.
\label{eq:first}
\end{equation}
or, equivalently 
\begin{equation}
H=\sqrt{\frac{h^2+N_j^2}{2}}\,,\qquad M=\frac{2}{\pi}\,\arctan\left(\frac{N_j}{h}\right)\;.
\label{eq:second}
\end{equation}
Obviously, the two-dimensional index $(H,M)$ is defined such that $0<H\leq N$ and $0<M\leq 1$. This representation separates the extensive contribution characterizing the volume of quality output, from an intensive characterization of its multidisciplinarity: small values of the argument $M$ indicates strong specialization, while $M$ close to one indicates  thematic dispersion with rather low impact. Fig.~\ref{fig:example2} reproduces the same set of data as Fig.~\ref{fig:example1} in these two variables. Interestingly, while $H$ is widely distributed, the average of $M$ discriminates more sharply between the different \enquote{Laboratoires}. One can even identify that the overlap between LPT and LPS is ensured by LPT researchers working in statistical physics. The overlap between data of LPA and LPS is due to the fact that LPA benefits from large publication material due the applied part of its research area (readership extent), while LPS researchers often work in different fields (multidisciplinarity). Clearly the $M$-index differentiates between communities and highlights readership diversity and research multidisciplinarity. The two indices together allow for a better qualification of research achievement compared to a single evaluator. Moreover, this alternative representation increases the level of independance between the two components, with a coefficient of correlation of 0.7 for $h$ and $N_j$ and $0.14$ for $H$ and $M$ in our dataset.
 
\section*{Discussion}

The use of the $h$-index to compare scientists from different research areas is a quite difficult task due to the inherent differences among different research fields. Although it is not an exclusive problem of the $h$-index, there have been several different efforts in the literature to supersede it. Interestingly, our work quantitatively demonstrates a bias associated with a citation-only index, namely its inability to capture research diversity. This is in favour to define $N_j$ as an additional index to supplement the $h$-index. However, the representation using the two-dimensional index $[H,M]$ is preferable since it yields weakly correlated components. Furthermore, while $M$ is bounded by 1 and separates finely the different disciplines, $H$, which is bounded by the number of papers, balances impact and diversity. In the following, we discuss questions that this two-dimensional index may rise and propose refinements to tackle its possible drawbacks.

\begin{itemize}
\item There are generalist journals that encompass different research areas and that could be downgraded by the proposed index. In order to overcome this difficulty, one could for example modify $N_j$ to categorize the paper using its research topic. Unfortunately, there is no uniform nomenclature between journal publishers (PACS numbers, keywords, subject classification,...). To do that, one would need a standard classification of research topics that is equivalent to Digital Object Identifier (DOI) system for publishers.

\item Let's try to compare two individuals with the same $H$. In an extreme case, an individual with $h=1$ and $N_j=N_0$, though carrying out multidisciplinary research, is not very successful since his research is not followed up by others. On the other side of the spectrum, a scientist with $h=N_0$ and $N_j=1$, though very exclusive in his publication choices, can be considered much more successful due to his citation rate. Hence having $M\approx1$ is never a good thing, while having $M\approx0$ can be of interest in some situations. That is why we claim that it would be ideal to have $h\approx N_j$, then $M\approx 1/2$, thus balancing recognition and curiosity.

\item The choice of the  \enquote{perfect} $M$ would of course strongly depend on the kind of target one aims at, but it lifts a degeneracy and provides deciders with an additional lever. The existence of a two-dimensional index refines the judgement depending of what one looks for. For example, hiring a researcher for a specific research position or for his teaching abilities is not the same, the former would require a smaller $M$ than the latter. Furthermore, evaluating an institution through citation metrics only is not fair because interdisciplinary research is an indisputable quality of an institution as it illustrates its local and international attractiveness and the diversity of education offered to students.


\end{itemize}

We claim that it is too limiting of an approach to categorize individuals along a single axis. Additional and independent indicators should complement such an approach. Our main line of axiomatics is that, independently of how \enquote{good or bad}  a researcher is ranked through the $h$-index, he may actually exhibit alternative qualities, like being a specialist, or conversely foraging results in various scientific areas. We claim that the two-dimensional index we introduced in Eq.~(\ref{eq:second}) is orthogonal to the usually accepted notion of \enquote{good or bad}.
Finally, we hope our proposition will trigger more bibliometric studies. Full-scale analysis of the $(H,M)$-index using different databases is an interesting question that will test its robustness and ease of computation. Before being adopted, it is necessary to apply it to real evaluation problems and situations: for example, additional studies comparing institutions should be developed. Nevertheless, we believe that combining the two numbers will refine degenerate situations and balance between impact and scope. Eventually, adding indices as legitimate as the $h$-index that quantify different facets of research achievements limit the misuse of the $h$-index.


\end{document}